\documentstyle[preprint,prb,aps]{revtex}
\begin{document}
\draft
\title{Comparison between the Torquato-Rintoul theory of the
interface effect in composite media and elementary results}
\author{Liang Fu and Lorenzo Resca}
\address{Department of Physics, the Catholic University of America,
Washington, D.C. 20064}
\maketitle
\begin{abstract}

We show that the interface effect on the properties of composite
media recently proposed by Torquato and Rintoul (TR) [Phys. Rev.
Lett. {\bf 75}, 4067 (1995)] is in fact
elementary, and follows directly from taking the limit in the
dipolar polarizability of a coated sphere: the TR ``critical
values'' are simply those that make the dipolar polarizability
vanish.  Furthermore, the new bounds developed by TR either
coincide with the Clausius-Mossotti (CM) relation or provide poor
estimates.  Finally, we show that the new bounds of TR do not
agree particularly well with the original experimental data that
they quote.

\end{abstract}

In a recent Letter,\cite{Torquato} Torquato and Rintoul (TR)
develop new bounds on the effective thermal
conductivity of a composite system with coated spheres for two
limiting cases, which consist in letting the thickness of the
coating layer vanish, and the conductivity of the coating layer
approach zero (resistance case) or infinity (conductance case), as
to maintain finite a given parameter.  They find that there are
``critical values'' for this parameter at which the inclusions are
``effectively hidden''.  They claim that ``the new bounds give
remarkably accurate predictions of the thermal conductivity''.

Alternatively, one can consider directly the dipolar polarizability
of a sphere with an outer radius $a$ and a coating layer of
thickness $\delta$, which is given by
\begin{eqnarray}
\gamma = {{(\sigma_s-\sigma_1)(\sigma_2+2\sigma_s)
+(\sigma_2-\sigma_s)(\sigma_1+2\sigma_s)(1-\delta /a)^3}
\over{(\sigma_2+2\sigma_s)(\sigma_s+2\sigma_1)
+2(\sigma_2-\sigma_s)(\sigma_s-\sigma_1)(1-\delta /a)^3}}a^3,
\end{eqnarray}
where $\sigma_2$, $\sigma_s$ and $\sigma_1$ refer to either thermal
or dielectric conductivities of the core, coating shell, and host
medium, respectively.  Taking the limits according to the
prescriptions of Eqs. (1) and (2) of Ref. \onlinecite{Torquato},
one obtains
\begin{eqnarray} 
\gamma ={{\alpha -1-R}\over{\alpha +2+2R}}a^3,
R=\lim_{{\delta\to 0}\atop{\sigma_s\to 0}}
\delta\sigma_2 /(a\sigma_s),
\end{eqnarray}
in case 1, and
\begin{eqnarray} 
\gamma ={{\alpha
-1+2C}\over{\alpha +2+2C}}a^3, C=\lim_{{\delta\to
0}\atop{\sigma_s\to \infty}} \delta \sigma_s/(a\sigma_1),
\end{eqnarray}
in case 2, having set $\alpha =\sigma_2/\sigma_1$.  These limits
may slightly simplify the expressions of the polarizability, but
they do not provide any particular gain, nor do they have any
special physical significance.  The procedure is just like that of
obtaining an ideal dipole as the limit of two opposite charges
becoming infinite while their separation vanishes. 

Now, if $R=\alpha -1$ in Eq. (2), or $C=(1-\alpha )/2$ in Eq. (3),
the polarizability  vanishes.  Hence, the TR result that the
inclusions are ``effectively hidden'' simply amounts to the
statement that for such ``critical values'' of $R$ or $C$ the
inclusions have no dipolar polarizability.

Using the polarizabilities given in our Eqs. (2) and (3), we have
thoroughly compared the results of the Clausius-Mossotti (CM)
relation with the bounds given in Eqs. (5)-(10) and (12)-(17) of
Ref. \onlinecite{Torquato}, for all values of the parameters
$\alpha$, $R$, and $C$.  We have found that, for all volume
fractions, the CM relation virtually coincides with the lower
(upper) bound in the resistance (conductance) case below the
critical point, and crosses over to coincide with the other bound
above the critical point.  Exceptions occur exclusively above the
critical point and for very large contrasts between core and
coating conductivities, in which case it is the TR bounds that
become ineffective.  On the other hand, there is a surprisingly
large range of conductivities of the constituents where both the
upper and the lower bound virtually coincide with the CM
relation, which would imply that the latter becomes essentially
exact.

We show in Figs. 1-2 (Figs. 3-4) typical examples of resistance and
conductance cases, below (above) the corresponding critical points.
All the cases in all the figures of Ref. \onlinecite{Torquato}
follow precisely this pattern.  The only exceptions to such
pattern occur above the critical point for $R>\alpha \gg 1$
(resistance case) and for $\alpha \ll 1$ and $C>(1-\alpha )/2$
(conductance case).  We show in Fig. 5 and 6 such exceptional
cases. However, it is clear that the cause of such exceptions 
is that the corresponding bounds of TR become increasingly
ineffective. Physically, $\sigma_e$ is expected to decrease as $R$
increases (see Fig. 3), whereas the upper bound in Fig. 5 does just
the opposite.  The same situation occurs in Fig. 6, with regard to
the lower bound.

All Figs. 1-6 refer to a volume fraction of 0.6, which is just
about the highest that can be achieved, compatibly with a uniform
distribution of spheres.  Hence, we have tested the CM relation
with the TR bounds precisely where CM is
known to be most inaccurate.  The result is that, except for
large-contrast situations where the TR bounds are ineffective, the
CM relation essentially provides a ``rigorous'' bound.
This is a surprising result, since we know that CM represents just
the mean-field approximation,\cite{Paper1,Paper3} and large
deviations from CM due to fluctuations have been shown to occur
both theoretically and by computer simulations.\cite{Felderhof} 
The TR bounds would then indicate that in each case the corrections
to the CM relation due to fluctuations occur only in one direction.

We have also found that, for all the three types of interfaces,
namely, resistance ($R>0$), conductance ($C>0$), and perfect
interface ($R=C=0$), there exists a surprisingly wide range of
conductivities of the constituents within which both the
upper and the lower bound virtually coincide with the CM relation. 
We illustrates in Fig. 7 the case of a perfect interface, i.e.,
uniform spheres with no coating, for $\alpha$ values between 0.25
and 3.5.  For such a wide range, where the effective conductivity
more than quadruples, the maximum error of the CM relation,
according to the TR bounds, is only about one percent.  This
applies again to all volume fractions, hence implies that in such
a wide range fluctuations beyond mean-field theory can hardly
occur.  We are not aware of any independent confirmation or
explanation of such a remarkable effect.  

Finally, we have carefully examined the only figure in
Ref. \onlinecite{Torquato} where TR compare their lower bounds with
experimental data, such bounds being virtually indistinguishable
from the CM relation.  Unfortunately, we have found that the curves
of Fig. 1 of Ref. \onlinecite{Torquato} in fact do {\it not} agree
particularly well with the original experimental data quoted in
Ref. \onlinecite{Torquato}.  We show in our Fig. 8 as diamonds and
crosses the original data,\cite{Araujo} and as squares
and triangles the data as portrayed in Fig. 1 of Ref.
\onlinecite{Torquato}.  Now, the original data points 
$\sigma_e$ for each temperature must be scaled by the same
constant $\sigma_1$, the thermal conductivity of the epoxy.  We
have determined $\sigma_1$ by the obvious criterion that the ratio
$\sigma_e /\sigma_1$ must approach unity at volume fraction $\Phi
=0$.  Anyway, whatever the scaling, it cannot account for the
substantial discrepancies between the original data in Ref.
\onlinecite{Araujo} and those portrayed in Fig. 1 of Ref. 
\onlinecite{Torquato}.  In fact, even the volume fractions of the
original data differ from those portrayed in Fig. 1 of Ref.
\onlinecite{Torquato}. Therefore, we
conclude that this experimental evidence does not support the
claim that ``the new bounds give remarkably accurate predictions of
the thermal conductivity''.\cite{Torquato}  Since
those bounds virtually coincide with the CM relation, we can also
conclude that a large body of experimental evidence in
general does not support such claim either.\cite{Bottcher}

Our work is supported in part by the U.S. Army Research Office
under contract No. DAAH04-93-G-0236.

\end{document}